\documentstyle[12pt,epsfig]{article}
\newlength{\dinwidth}
\newlength{\dinmargin}
\newcommand{\ba}{\begin{array}}
\newcommand{\ea}{\end{array}}
\newcommand{\be}{\begin{equation}}
\newcommand{\ee}{\end{equation}}
\newcommand{\bea}{\begin{eqnarray}}
\newcommand{\eea}{\end{eqnarray}}

\def\l{\lambda}

\def\d{{\rm d}}

\def\void{}
{\ifx\void\labelname\def\junk{\end{displaymath}}
\else\def\junk{\end{equation}}\fi\junk\labelmark\def\labelname{}}
{\ifx\void\labelname\def\junk{\end{array}\end{displaymath}}
\else\def\junk{\end{array}\right.\end{equation}}
\fi\junk\labelmark\def\labelname{}\def\junk{}
}
{\ifx\void\labelname\def\junk{\end{displaymath}}
\else\def\junk{\end{eqnarray}}\fi\junk\labelmark\def\labelname{}}

\def\l{\left}
\def\r{\right}
\def\w{\wedge}

\newfont{\sgcal}{eufm9}
\newfont{\smcal}{eusm9}
\newfont{\smbf}{msbm9}
\newfont{\gcal}{eufm10}
\newfont{\mcal}{eusm10}
\newfont{\mbf}{msbm10}
\newfont{\Gcal}{eufm10 scaled\magstep1}
\newfont{\Mcal}{eusm10 scaled\magstep1}
\newfont{\Mbf}{msbm10 scaled\magstep1}
\def\l{\left}
\def\r{\right}
\def\w{\wedge}

\begin{document}
\thispagestyle{empty}
\addtocounter{page}{-1}
\begin{flushright}
SNUST 01-0501\\
IHP/2001/19\\
{\tt hep-th/0106121}\\
\end{flushright}
\vspace*{1.3cm}
\centerline{\Large \bf Open Wilson Lines and Generalized Star Product}
\vspace*{0.4cm}
\centerline{\Large \bf in Noncommutative Scalar Field Theories~\footnote{
Work supported in part for SJR and JTY by BK-21 Initiative in Physics through 
(SNU-Project 2), KOSEF Interdisciplinary Research Grant 98-07-02-07-01-5, 
and KOSEF Leading Scientist Program 2000-1-11200-001-1, for HTS by KOSEF
Brain-Pool Program, and for YK by KRF Grant 2001-015-DP0082. }}
\vspace*{1cm} 
\centerline{\bf Youngjai Kiem ${}^a$, Soo-Jong Rey ${}^{b,c}$, 
Haru-Tada Sato ${}^a$, Jung-Tay Yee ${}^{b,c}$ }
\vspace*{0.5cm}
\centerline{\it BK21 Physics Research Division \& 
Institute of Basic Science}
\vspace*{0.25cm}
\centerline{\it Sungkyunkwan University, Suwon 440-746 \rm KOREA ${}^a$} 
\vspace*{0.4cm}
\centerline{\it School of Physics \& Center for Theoretical Physics}
\vspace*{0.25cm}
\centerline{\it Seoul National University, Seoul 151-747 \rm KOREA ${}^b$}
\vspace*{0.4cm}
\centerline{\it Centre Emil Borel, Institut Henri Poincar\'e}
\vspace*{0.25cm}
\centerline{\it 11, rue Pierre et Marie Curie, Paris F-75231 \rm FRANCE ${}^c$}
\vspace*{0.5cm}
\centerline{\tt ykiem, haru@newton.skku.ac.kr 
\hskip0.5cm sjrey@gravity.snu.ac.kr \hskip0.5cm jungtay@phya.snu.ac.kr}
\vspace*{1.5cm}
\centerline{\bf abstract}
\vspace*{0.5cm}
Open Wilson line operators and generalized star product have been studied 
extensively in noncommutative gauge theories. We show that they also show up 
in noncommutative scalar field theories as universal structures. We first 
point out that 
dipole picture of noncommutative geometry provides an intuitive argument for
robustness of the open Wilson lines and generalized star products therein.
We calculate one-loop effective action of noncommutative scalar field theory 
with cubic self-interaction and show explicitly that the generalized star 
products arise in the nonplanar part. It is shown that, at low-energy, large 
noncommutativity limit, the nonplanar part is expressible solely in terms of 
the {\sl scalar} open Wilson line operator and descendants. 
\vspace*{1cm}

\begin{flushleft}
\end{flushleft}

\baselineskip=18pt
\newpage

\section{Introduction and Summary}
\setcounter{equation}{0}
\indent
One of the most salient features of noncommutative field theories is that
physical excitations are described by \underline{`dipoles'} --- 
weakly interacting, nonlocal objects. Denote their 
center-of-mass momentum and dipole moment as ${\bf k}$ and $\Delta {\bf x}$, 
respectively. According to the `dipole' picture, originally developed in
\cite{dipole} and more recently reiterated in \cite{susskind}, 
the two are related each other:
\bea
\Delta {\bf x}^a = \theta^{ab} {\bf k}_b.
\label{relation}
\eea
Here, $\theta^{ab}$ denotes the noncommutativity parameter:
\bea
\left\{ {\bf x}^a, {\bf x}^b \right\}_\star = i \theta^{ab}
\label{nc}
\eea
in which $\left\{ \cdot \right\}$ refers to the Moyal commutator, defined
in terms of the $\star$-product:
\bea
\left\{ A(x_1) B(x_2) \right\}_\star
:= \exp \left( {i \over 2} \partial_1 \wedge \partial_2 \right) A(x_1) B(x_2)
\qquad
{\rm where}
\qquad
\partial_1 \wedge \partial_2 := \theta^{ab} \partial_1^a \partial_2^b.
\eea
Evidently, in the commutative limit, the dipoles shrink in size 
and represent pointlike excitations. 

In noncommutative {\sl gauge} theories, (part of) the gauge orbit is
the same as the translation along the noncommutative directions \cite{rey1, 
rey2, gross}. For example,
in noncommutative U(1) gauge theory, the gauge potential ${\bf A}_\mu(x)$
and the neutral scalar field ${\Phi}(x)$, both of which give rise to 
`dipoles', transform in `adjoint' representation:
\bea
\delta_\epsilon {\bf A}_\mu (x) &=& i \int {\d^2 {\bf k} \over (2 \pi)^2}
\, \widetilde{\epsilon}({\bf k})  
\Big[ \Big( {\bf A}_\mu ({\bf x} + \theta \cdot {\bf k}) - 
{\bf A}_\mu ({\bf x} - \theta \cdot {\bf k}) \Big) + i {\bf k}_\mu \Big]
e^{ i {\bf k} \cdot {\bf x}} \nonumber \\
\delta_\epsilon {\Phi}(x)
&=& i \int {\d^2 {\bf k} \over (2 \pi)^2} \, 
\widetilde{\epsilon}({\bf k}) \Big[ {\Phi} ({\bf x} + \theta \cdot {\bf k})
- {\Phi} ({\bf x} - \theta \cdot {\bf k}) \Big] e^{ i {\bf k} \cdot {\bf x}},
\label{gaugetransf}
\eea
where the infinitesimal gauge transformation parameter is denoted as
\bea
\epsilon({\bf x}) = \int {\d^2 {\bf k} \over (2 \pi)^2}
e^{ i {\bf k} \cdot {\bf x}} \, \widetilde{\epsilon}({\bf k}).
\nonumber
\eea 
For fields transforming in `adjoint' representations under the noncommutative
gauge group, it has been shown that the only physical observables are the
`open Wilson lines' \cite{kawai, rey1, rey2, gross}
along an open contour $C$:
\bea
W_{\bf k} [C] = {\cal P}_{\rm t} \int {\d^2} {\bf x} \,
\exp_\star \left[ i \int_0^1 \d t \, \dot{\bf y}(t) \cdot {\bf A}
(x + {\bf y}(t))
\right] \star e^{ i {\bf k} \cdot {\bf x}},
\label{openwilson}
\eea
and their descendant operators. The $\star$-product refers to the base point 
${\bf x}$ of the open contour $C$.
Despite being defined over an open contour,
the operator is gauge-invariant {\sl provided} the momentum ${\bf k}$ is
related to the geodesic distance ${\bf y}(1) - {\bf y}(0) := \Delta {\bf x}$
precisely by the `dipole relation', Eq.(\ref{relation}). In other words, 
in noncommutative gauge theory, the open Wilson lines (physical
observables) ought to obey the dipole relation Eq.(\ref{relation}) 
as an immediate consequence of the gauge invariance! 
For a straight Wilson line, expanding Eq.(\ref{openwilson}) in 
successive powers of 
the gauge field ${\bf A}_\mu$, it was observed \cite{mehenwise, liu3}
that generalized $\star$-product, $\star_n$, a structure discovered first
in \cite{garousi, liu}, emerge:
\bea
W_{\bf k} [C] = \int \d^2 {\bf x} \, 
\left[ 1 - \left( \partial \wedge {\bf A} \right) +
{1 \over 2!} \left(\partial \wedge {\bf A} \right)^2_{\star_2} 
+ \cdots \right] \star 
e^{ i {\bf k} \cdot {\bf x}}.
\label{wilsonline2}
\eea

Apparently, the generalized $\star_n$ products give rise to different 
algebraic structures from Moyal's $\star$-product. For instance, the 
first two, $\star_2, \star_3$ defined as: 
\bea
\left[A(x_1) B(x_2) \right]_{\star_2} 
&:=& { \sin \left({1 \over 2} \partial_1 \wedge \partial_2 \right) 
\over {1 \over 2} \partial_1 \wedge \partial_2} A(x_1) B(x_2) 
\nonumber \\
\left[ A(x_1) B(x_2) C(x_3) \right]_{\star_3}
&:=& \left[ {\sin \left( {1 \over 2} \partial_2 \wedge \partial_3 \right)
\over {1 \over 2}(\partial_1 + \partial_2) \wedge \partial_3 }
{\sin \left({1 \over 2} \partial_1 \wedge (\partial_2 +\partial_3) \right)
\over {1 \over 2} \partial_1 \wedge (\partial_2 +\partial_3)}
+(1 \leftrightarrow 2) \right] A(x_1) B(x_2) C(x_3)
\nonumber
\eea  
show that the $\star_n$'s are commutative but non-associative. 
Even though the definition of the open 
Wilson line is given in terms of path-ordered $\star$-product, its expansion 
in powers of the gauge potential involves the generalized $\star_n$-product 
at each $n$-th order. The complicated $\star_n$ products have arisen upon 
expansion in powers of the gauge potential, 
and are attributible again to dipole nature of the Wilson line and the gauge 
invariance thereof -- 
each term in Eq.(\ref{wilsonline2}) is {\sl not} gauge invariant, as
the gauge transformation Eq.(\ref{gaugetransf}) mixes terms involving 
different $\star_n$'s. Indeed, the generalized $\star_n$ 
products are not arbitrary but obey recursive identities:
\bea
i \left[\partial_x A(x) \wedge \partial_x B(x) \right]_{\star_2}
&=& \left\{ A(x), B(x) \right\}_{\star} \nonumber \\
i \partial_x \wedge \left[ A(x) B(x) \partial_x C(x) \right]_{\star_3}
&=& A(x) \star_2 \left\{ B(x), C(x) \right\}_\star 
+ B(x) \star_2 \left\{A(x), C(x) \right\}_\star. 
\nonumber
\eea
These recursive identities are crucial for ensuring gauge invariance 
of the power-series expanded open Wilson line operator, 
Eq.(\ref{wilsonline2}).

The open Wilson lines and the generalized $\star$-products 
therein have been studied extensively in literatures, largely
in the context of noncommutative {\sl gauge} theories. On the other hand, the 
dipole relation Eq.(\ref{relation}) ought to be a universal relation, 
applicable for {\sl any} theories defined over noncommuative spacetime.
The aforementioned remark that noncommutative gauge invariance plays a 
prominent role in both deriving the relation Eq.(\ref{relation}) and 
the recursive relation among the generalized $\star$-products then
raises an immediate question. Are these structures present also in 
noncommutative field theories, in which neither gauge invariance nor
gauge field exists?

In this paper, we show that the answer to the above question
is affirmatively \underline{yes} by studying $d$-dimensional 
massive $\lambda \left[{\Phi}^3 \right]_\star$ theory. 
We compute the one-loop effective action and find that nonplanar 
contributions are expressible in terms of the generalized $\star$-product. 
Most significantly, at low-energy and large noncommutativity limit,
we show that the nonplanar part of the one-loop effective action can 
be resummed into a remarkably simple expression
\bea
\Gamma_{\rm np}[{\Phi}]
= {\hbar \over 2} \int \!\! {\d^d k \over (2 \pi)^d} 
W_{\bf k}[{\Phi}] \, \widetilde{{\cal K}_{-{d \over 2}} }
\left({\bf k} \circ {\bf k} \right) \, W_{- \bf k}[\Phi],
\label{final}
\eea
where $W_{\bf k}[\Phi]$ denotes the first descent of the scalar Wilson line 
operators:
\bea
W_{\bf k}[\Phi] &:=& {\cal P}_{\rm t} \!\! \int \!\! \d^2 {\bf x}\,
\exp \left( - g \int_0^1 \d t \vert \dot{\bf y}(t) \vert \,
\Phi(x + {\bf y}(t))  \right) \star e^{ i {\bf k} \cdot {\bf x}}
\label{scalarwilson} \\
(\Phi W)_{\bf k} [\Phi]
&:=& {\cal P}_{\rm t} \!\! \int \!\! \d^2 {\bf x} 
\left[ \Big( \int_0^1 \d t \vert \dot{\bf y}(t) \vert \, 
\Phi(x + {\bf y}(t)) \Big) 
\exp \left( - g\int_0^1 \d t \vert\dot{\bf y}(t)\vert\,
             \Phi(x + {\bf y}(t) \right) \right] 
\star e^{ i {\bf k}\cdot{\bf x}}
\nonumber \\
&=&
\left( - {\partial \over \partial g} \right) W_{\bf k}[\Phi]
\nonumber\\
\cdots \nonumber \\
(\Phi^n W)_{\bf k}[\Phi]
&:=& \left( - {\partial \over \partial g} \right)^n W_{\bf k}[\Phi],
\qquad \qquad \Big( g := { \lambda \over 4 m} \Big)
\label{scalaropenwilson}
\eea
Here, 
$\widetilde{{\cal K}_{-{d \over 2}}}$ denotes a (Fourier-transformed) kernel
and ${\bf k} \circ {\bf k} := \left( \theta \cdot {\bf k} \right)^2$, etc.
We trust that the above result bears considerable implications to the
noncommutative solitons studied in \cite{GMS, harvey} and to the UV/IR mixing 
discovered in \cite{seiberg}, and will report the details elsewhere.
\section{Scalar Open Wilson Lines as Dipoles}
\subsection{Dipole Relation for Scalar Fields}
We begin with a proof that the dipole relation Eq.(\ref{relation}) holds 
for scalar fields as well, wherein no gauge invariance is present. Recall 
that, in noncommutative spacetime, energy-momentum is
a good quantum number, as the noncommutative geometry is invariant under
translation along the noncommutative directions. As such, consider the
following set of operators, so-called Parisi operators \cite{parisi}:
\bea
{\cal O}_n (x_1, \cdots, x_n; {\bf k}) 
= \int d^2 {\bf z} \, \Phi_1 (x_1 + {\bf z}) \star \Phi_2 (x_2 + {\bf z})
\star \cdots \star \Phi_n (x_n + {\bf z}) \star 
e^{ i {\bf k} \cdot {\bf x}},
\nonumber
\eea
viz. Fourier-transform of a string of elementary fields, $\Phi_k(x)$
$(k = 1, 2, \cdots)$. 
Consider the one-point function:
\bea
G_1 (x, {\bf k}) &=& \left< {\cal O}_2 (x, {\bf k}) \right>
\nonumber \\
&=& \left< \int d^2{\bf z} \, \Phi({\bf z}) \star\Phi(x + {\bf z}) \star
e^{ i {\bf k} \cdot {\bf x}} \right>.
\nonumber
\eea
In terms of Fourier decomposition of the scalar field:
\bea
\Phi(x) = \int {d^2 {\bf k} \over (2 \pi)^2} \,
e^{ i {\bf k} \cdot {\bf x}} \widetilde{\Phi}({\bf k}),
\nonumber
\eea
one obtains that
\bea
G_1 (x, {\bf k}) = \int {d^2 {\bf l} \over (2 \pi)^2}
\widetilde{\Phi}({\bf l}) \widetilde{\Phi}(-{\bf l} + {\bf k})
\exp\left[ i {\bf l} \cdot 
\left(x + {1 \over 2} \theta \cdot {\bf k} \right)\right].
\nonumber
\eea
Consider `wave-packet' of the scalar particle, $\Phi({\bf z}) = 
\Phi_0 \delta^{(2)} ({\bf z})$ and $\Phi(x +{\bf z}) = 
\Phi_0 \delta^{(2)} (x + {\bf z})$ so that $\widetilde{\Phi}({\bf l})
= \Phi_0 \exp(i {\bf l} \cdot x)$. From the above equation, one
then finds that 
\bea
G_1 (x, {\bf k}) = \Phi_0^2 
\delta^{(2)} \left( x +{1 \over 2} \theta \cdot {\bf k} \right).
\nonumber
\eea
Thus, one finds that the stationary point of the correlator is given 
by 
\bea
\Delta x^a \, \sim \, \theta^{ab} {\bf k}_b 
\label{scalarrelation}
\eea
and hence precisely by the `dipole relation', Eq.(\ref{relation}).

\subsection{Scalar Open Wilson Lines and Generalized Star Products}
First, as in the case of the {\sl gauge} open Wilson lines, 
Eq.(\ref{openwilson}), we will show that power-series expansion of the 
{\sl scalar} open Wilson line operators Eq.(\ref{scalaropenwilson}) gives
rise to generalized $\star_n$ products, and take this as the definition
of the products for `adjoint' scalar fields. 

Begin with a remark concerning symmetry of the open Wilson line operators. 
The open Wilson line operator Eq.(\ref{openwilson}) possesses 
reparametrization invariance:
\bea
t  \rightarrow \tau(t)  \qquad {\rm such} \,\, {\rm that}
 \qquad \tau(0)=0, \tau(1) = 1 \quad {\rm and} \quad \dot{\tau}(t) > 0,
\label{repara}
\eea
as the gauge-element $\d {\bf y}(t) \cdot {\bf A}(x + {\bf y})$ is invariant 
under the reparametrization, Eq.(\ref{repara}):
\bea
\d t \, \dot {\bf y} (t) \cdot {\bf A}(x + {\bf y}(t))  = 
\left( \d \tau \vert \dot{\tau}(t) \vert^{-1} \right) \left(\dot{\tau}(t) \,
\dot{\bf y} (\tau) \right) \cdot {\bf A}(x + {\bf y}(\tau))
= \d \tau \, \dot{\bf y}(\tau) \cdot {\bf A}(x +{\bf y}(\tau)).
\nonumber
\eea
Likewise, the {\sl scalar} open Wilson line operator 
Eq.(\ref{scalaropenwilson}) posseses reparametrization invariance, as 
the scalar-element $\vert \d {\bf y}(t) \vert \Phi(x +{\bf y})$ is 
invariant under the reparametrization, Eq.(\ref{repara}):
\bea
\d t \vert \dot{\bf y}(t) \vert \Phi(x + {\bf y}(t))
= \left( \d \tau \vert \dot{\tau}(t) \vert^{-1} \right)
\left( \vert \dot{\tau}(t) \, \dot{\bf y}(\tau) \vert \right)
\Phi(x + {\bf y}(\tau))
= \d \tau \vert \dot {\bf y}(\tau) \vert \Phi(x + {\bf y}(\tau)).
\nonumber
\eea
In fact, the reparametrization invariance, along with rotational invariance
on the noncommutative subspace and reduced Lorentz invariance on the complement
spacetime, puts a powerful constraint enough to determine uniquely
structure of the scalar open Wilson line operator, Eq.(\ref{scalaropenwilson}),
much as for that of the gauge open Wilson line operator, Eq.(\ref{openwilson}).

For illustration, we will examine a simplified form of the scalar open
Wilson line with an insertion of a local operator ${\cal O}$ at a location
${\bf R}$ on the Wilson line:
\bea
({\cal O}_{\bf R} W)_{\rm k} [\Phi]
:= {\cal P}_{\rm t}
\int \d^2 {\bf x} \, {\cal O}(x + {\bf R}) \star
\exp \left( - g \int_0^1 \d t \vert \dot{\bf y}(t) \vert 
\Phi(x + {\bf y}(t) ) \right) \star e^{ i {\bf k} \cdot
{\bf x}}.
\nonumber
\eea
Take a {\sl straight} Wilson line:
\bea
{\bf y}(t) = {\bf L} t \qquad
{\rm where} \qquad {\bf L}^a = \theta^{ab} {\bf k}_b := \left( \theta
\cdot {\bf k} \right)^a, \quad
L := \vert {\bf L} \vert,
\nonumber
\eea 
corresponding to a uniform distribution of the momentum ${\bf k}$ along the 
Wilson line. As the path-ordering progresses to the right with increasing 
$t$, power-series expansion in $g L \Phi$ yields:
\bea
({\cal O}_{\bf R} W)_{\rm k}[\Phi]
&=& \int \d^2 {\bf x} {\cal O}({\bf x} + {\bf R}) \star
e^{ i {\bf k} \cdot {\bf x}}
\nonumber \\
&+& (- gL) \int \d^2 {\bf x} \int_0^1 \d t {\cal O}(x + {\bf R})
\star \Phi(x + {\bf L} t) \star e^{ i {\bf k} \cdot {\bf x}}
\nonumber \\
&+& (- gL)^2 \int \d^2 {\bf x} \int_0^1 \d t_1 \int_{t_1}^1 \d t_2 
{\cal O}({\bf x} + {\bf R}) \star \Phi(x + {\bf L} t_1)
\star \Phi(x + {\bf L} t_2 ) \star e^{i {\bf k} \cdot {\bf x}}
\nonumber \\
&+& \cdots.
\nonumber
\eea
Each term can be evaluated, for instance, by Fourier-transforming 
${\cal O}$ and $\Phi$'s:
\bea
{\cal O}(x) = \int {\d^2 {\bf k} \over (2 \pi)^2}
\widetilde{O}({\bf k}) \, { T}_{\bf k}, \quad
\Phi(x) = \int {\d^2 {\bf k} \over (2 \pi)^2}
\widetilde{\Phi}({\bf k}) \, {T}_{\bf k} \qquad
{\rm where} \qquad
{T}_{\bf k} = e^{ i {\bf k} \cdot {\bf x}},
\nonumber
\eea
taking the $\star$-products of the translation generators $T_{\bf k}$'s
\bea
{T}_{\bf k} \star {T}_{\bf l} = e^{ {i \over 2} {\bf k} \wedge {\bf l}}
\, {T}_{{\bf k} + {\bf l}},
\nonumber
\eea
and then evaluating
the parametric $t_1, t_2, \cdots$ integrals. Fourier-transforming 
back to the configuration space, after straightforward calculations, 
one obtains
\bea
({\cal O}_{\bf R} W)_{\rm k}[\Phi]
&=& \int \d^2 {\bf x} \, e^{ i {\bf k} \cdot {\bf x}} {\cal O}(x + {\bf R})
\nonumber \\
&+& (- gL)  \int \d^2 {\bf x} \, e^{ i {\bf k} \cdot {\bf x}} 
\left[ {\cal O}(x + {\bf R}) \Phi(x)\right]_{\star_2}
\nonumber \\
&+& {1 \over 2!} (-gL)^2  \int \d^2 {\bf x} \, e^{ i {\bf k} \cdot {\bf x}}
\left[{\cal O}(x + {\bf R}) \Phi(x) \Phi(x)
\right]_{\star_3} 
\nonumber \\
&+& \cdots,
\label{taylor}
\eea
where the products involved, $\star_2, \star_3, \cdots$, are precisely the
same generalized star products as those apparing in the {\sl gauge} open 
Wilson line operators.  

In fact, for the straight open Wilson lines, one can take each term of
the Taylor expansion in Eq.(\ref{taylor}) as the definition of the
generalized $\star_n$ products. Denote the exponent of the scalar open
Wilson line operator as:
\bea
 \Phi_\ell \left(x,{\bf k} \right) \, := \, 
\int_0^1 \d t_\ell \vert \theta \cdot {\bf k} \vert 
\Phi \left(x+ \theta \cdot {\bf k} t_\ell \right).
\nonumber
\eea
We have taken the open Wilson lines as straight ones. Then, analogous to
\cite{liu}, the generalized $\star_n$ products are given by:
\bea
 \left[ \Phi\star_n \Phi \right]_{\bf k} &:=& 
{\cal P}_{\rm t} \int \d^d x \,  
\Phi_1 \left(x,{\bf k} \right) \star \Phi_2 \left(x, {\bf k} \right) 
\cdots \Phi_n \left(x , {\bf k} \right) \star e^{i{\bf k} \cdot {\bf x}} 
 \nonumber \\
 &= & \int \d^d x \, \Phi^{\star n}(x) \star e^{i{\bf k} \cdot
{\bf x}}.
\label{generalizedstar} 
\eea
\section{ One-Loop Effective Action in $ \lambda [\Phi^3]_\star$ Theory}
Given the above results, we assert that the {\sl scalar} open Wilson line 
operators play a central role in noncommutative field theories. To support
our assertion, we now show that, for the case of 
$\lambda [\Phi^3]_\star$ scalar field theory,
the effective action is expressible entirely in terms of the open Wilson line
operators and nothing else. Begin with the classical action 
\bea
S_{\rm NC} = \int {\d^d} x \, \left[{1 \over 2} (\partial_\mu \Phi)^2 
- {1 \over 2} m^2 \Phi^2  - {\lambda \over 3!} \Phi \star \Phi \star \Phi
\right].
\label{action}
\eea
At one-loop order, the effective action can be obtained via the background 
field method \footnote{Technical details of the following results will be 
reported in a separate paper \cite{forthcoming}.} --- expand the scalar field 
around a classical configuration, $\Phi = \Phi_0 + \varphi$:
\bea
S_{\rm NC} = \int \d^d x \, {1 \over 2} 
\varphi(x) \star
\left[ - \partial_x^2 - m^2 - \lambda \Phi_0(x) \right] \star \varphi(x)
\eea
and then integrate out the quantum flucutation, $\varphi$. Fourier transform
yields the interaction vertex of the form:
\bea
iS_{\rm int.}
&=&-i {\lambda \over 2 \cdot 2} \int \prod_{a=1}^3 {\d^d k_a \over (2 \pi)^d} \,
\widetilde{\varphi}(k_1) \widetilde{\Phi_0}(k_2) \widetilde{\varphi}(k_3) \,
\left[e^{ - {i \over 2} k_1 \wedge k_2} 
+ e^{ + {i \over 2} k_1 \wedge k_2} \right]
(2 \pi)^d \delta^{(d)} (k_1 + k_2 + k_3)
\nonumber \\
&:=& \Big( {\bf U} + {\bf T} \Big),
\label{vertex}
\eea
under the rule that $\varphi$'s are Wick-contracted in the order they appear
in the Taylor expansion of $\exp \left(i S_{\rm int} \right)$. The {\bf U} and 
{\bf T} parts refer to the untwisted and the twisted vertex interactions, 
respectively. In Eq.(\ref{vertex}), relative sign between 
${\bf T}$ and ${\bf U}$ is +, opposite to those for gauge fields. This can be
traced to the fact that, under `time-reversal' $t \rightarrow (1 - t)$, the
exponents in gauge and scalar open Wilson lines, Eq.(\ref{openwilson})
and Eq.(\ref{scalaropenwilson}), transform as even and odd, respectively.
The N-point correlation function is given by 
$(iS_{\rm int})^{\rm N} \slash {\rm N}!$ term in 
the Taylor expansion and, as shown by Filk \cite{filk}, 
consists of planar and non-planar contributions. In the binomial expansion
of $(i S_{\rm int})^{\rm N} = ({\bf U} + {\bf T})^{\rm N}$, 
the two terms, $[{\bf U}]^{\rm N}$ 
and $[{\bf T}]^{\rm N}$, 
yield the planar contribution, while the rest comprises
the nonplanar contribution. Evaluation of the N-point correlation function
is straightforward. Summing over N, the one-loop effective action is
given as:
\bea
\Gamma [\Phi_0]
&:=& -{\hbar \over 2} \ln \det 
\left[ - \partial_x^2 - m^2 - \lambda 
\Phi_0(x) \star - i \epsilon \right]_\star
\nonumber \\
&=& \sum_{\rm N=1}^\infty 
 \int \prod_{\ell = 1}^{\rm N} {\d^d q_\ell 
\over (2 \pi)^d} \, \widetilde{\Phi_0}(q_1) \cdots \widetilde{\Phi_0}
(q_N) \, {\bf \Gamma}^{\rm (N)} \left[q_1, \cdots, q_{\rm N} \right].
\nonumber
\eea
Here, the N-point correlation function takes schematically the following form: 
\bea
{\bf \Gamma}^{\rm (N)} \left[q_1, \cdots, q_{\rm N} \right]
=  (2 \pi)^d \delta^{(d)} (q_1 + \cdots + q_{\rm N}) \, 
{\cal S}_{\rm N} \cdot {\cal A}(q_1, \cdots, q_{\rm N}) \cdot 
{\cal B} (q_1, \cdots, q_{\rm N}), 
\label{form}
\eea
where ${\cal S}_{\rm N}$ refers to the combinatorics factor, 
${\cal A}$ is the Feynman loop integral, and 
${\cal B}$ consists of the noncommutative phase factors. 

\subsection{Effective Action: Non-planar Part}
We are primarily interested in the nonplanar part. After a straightforward
algebra, we have obtained the nonplanar part as a double sum involving
the generalized star products:
\bea
\Gamma_{\rm np} = \hbar
\sum_{\rm N=2}^{\infty} \l(- { \lambda \over 4}\r)^{\rm N} 
  {1 \over {\rm N}!}
\int \, \d^d x \,\,  \l( 
{1 \over 2} \sum_{n=1}^{\rm N-1} {{\rm N} \choose n} 
\Big[\Phi_0\star_n\Phi_0 \Big](x) \, \cdot \,
{\cal K}_{{\rm N} - {d \over 2}} \, \cdot \,
\Big[\Phi_0 \star_{{\rm N}-n} \Phi_0 \Big](x) \r). 
\nonumber
\eea
Here, $[\Phi_0 \star_n \Phi_0] (x)$ refers to the generalized $\star_n$ 
products of $\Phi_0$'s and the combinatoric factor ${1 \over 2} {{\rm N} 
\choose n}$ originates from the binomial expansion of $({\bf U} + {\bf T})^{\rm N}$ modulo an inversion symmetry, corresponding to Hermitian conjugation.
The kernel ${\cal K}_n$ is defined as
\bea
{\cal K}_n := {\cal K}_n \left( - \partial_x \circ \partial_x \right) 
\qquad
{\rm where} 
\qquad
{\cal K}_n \left( z^2 \right) 
= 2 \left( {1 \over 2 \pi} \right)^{d \over 2}
\left( {\vert z \vert \over m} \right)^{n}
{\bf K}_n \Big( m \vert z \vert \Big)
\eea
in terms of the modified Bessel functions, ${\bf K}_n$.

To proceed further, we will be taking the low-energy, large noncommutativity 
limit:
\bea
q_\ell \sim \epsilon, \qquad
{\rm Pf} \theta \sim {1 \over \epsilon^2} \qquad {\rm as}
\qquad \epsilon \rightarrow 0
\label{limit1}
\eea
so that
\bea
q_\ell \cdot q_m \sim {\cal O} (\epsilon^{+2}) \rightarrow 0, 
\qquad
q_\ell \wedge q_m \rightarrow {\cal O}(1), 
\qquad
q_\ell \circ q_m \sim {\cal O}(\epsilon^{-2}) \rightarrow \infty.
\label{limit2}
\eea
In this limit, the modified Bessel function ${\bf K}_n$ exhibits 
the following asymptotic behavior:
\bea
 {\bf K}_n \Big(m z \Big) \rightarrow \sqrt{ {\pi \over 2 m z}} e^{-m \vert
z \vert} 
\left[ 1+ {\cal O} \l({1 \over m \vert z \vert}\r) \right].
\eea
Most importantly, the asymptotic behavior is {\sl independent} of 
the index $n$. Hence, in the low-energy limit, the Fourier-transformed 
kernels, $\widetilde{{\cal K}_n}$'s obey the following recursive relation:
\bea
\widetilde{{\cal K}^{}}_{n+1} \left( {\bf k} \circ {\bf k} \right) = 
\left( \vert\theta \cdot {\bf k} \vert \over m \right)
\, \widetilde{{\cal K}^{}}_n \left( {\bf k} \circ {\bf k} \right),
\eea
viz.
\bea
\widetilde{{\cal K}_n} \left({\bf k} \circ {\bf k} \right)
= \left( { \vert \theta \cdot {\bf k} \vert \over m} \right)^n 
\, \widetilde{{\cal Q}^{}} \left({\bf k} \circ {\bf k} \right).
\eea
Here, the kernel $\widetilde{{\cal Q}_{}}$ is given by:
\bea
\widetilde{{\cal Q}_{}} \left({\bf k} \circ {\bf k} \right)
= \left( 2 \pi \right)^{(1 - d)/2} \left\vert {1 \over m \, \theta \cdot {\bf k}} \right\vert^{1 \over 2}  \exp \Big(- m \vert \theta \cdot {\bf k}  \vert 
\Big).
\eea
Note that, in power series expansion of the effective action, natural 
expansion parameter is $\vert \theta \cdot {\bf k} \vert$ 
\footnote{As will be shown momentarily, this implies that manifestly 
reprametrization invariant open Wilson 
line operator Eq.(\ref{scalaropenwilson}) is more fundamental than those
defined in \cite{okawaooguri} and utilized in \cite{dastrivedi, mukhi, 
liumichelson}.  }.

Thus, the nonplanar one loop effective action in momentum space 
is expressible as:
\bea
 \Gamma_{\rm np}[\Phi] &=&{\hbar \over 2} \int {\d^d k \over (2\pi)^d} 
\widetilde{\cal K}_{-{d \over 2}} \left({\bf k} \circ {\bf k} \right)
\sum_{{\rm N}=2}^{\infty} \sum_{n=1}^{\rm N-1}
\left(- {\lambda \over 4 m }\right)^{\rm N}
\nonumber \\
&\times & \left( {1 \over n!} \vert \theta \cdot {\bf k} \vert^n 
\left[ \widetilde{\Phi} \star_n \widetilde{\Phi} \right]_{\bf k} \right)
\left( {1 \over ({\rm N} - n)!} \vert\theta \cdot {\bf k}\vert^{{\rm N} - n} 
\left[ \widetilde{\Phi} \star_{{\rm N} - n} \widetilde{\Phi} \right]_{-\bf k}
\right).
\label{seriesform}
\eea
Utilizing the relation between the generalized $\star_n$ products and the
{\sl scalar} open Wilson line operators, as elaborated in section 2,
the nonplanar one-loop effective action can be summed up into a 
remarkably simple closed form. Denote the rescaled coupling parameter as
$g := \lambda / 4m$ (see Eq.(\ref{scalaropenwilson})). Exploiting the 
exchange symmetry $n \leftrightarrow ({\rm N} - n)$, one
can rearrange the summations into {\sl decoupled} ones over $n$ and 
$({\rm N}-n)$. Moreover, because of $[\widetilde{\Phi} \star_0 
\widetilde{\Phi}]_{\bf k} = (2 \pi)^d \delta^{(d)} (k)$, 
the summations can be extended 
to $n=0$, (N$-n)=0$ terms, as they yield identically vanishing contribution
after $k$-integration \footnote{ We assume that the noncommutativity is 
turned only on two-dimensional subspace.}. One finally obtains: 
\bea
\Gamma_{\rm np}[\Phi] = 
   {\hbar \over 2} 
\int {\d^d k \over (2 \pi)^d}\,
 W_{\bf k}[\Phi] \cdot \widetilde{{\cal K}_{-{d \over 2}}}
\left({\bf k} \circ {\bf k} \right) \cdot W_{-{\bf k}}[\Phi],
\nonumber
\eea
yielding precisely the aforementioned result, Eq.(\ref{final}).

\subsection{Explicit Calculations}
To convince the readers that the expression Eq.(\ref{final}) is
indeed correct, we will evaluate below the simplest yet nontrivial 
correlation functions: N= 3, 4. Utilizing the factorization property
\cite{factorization}, begin with nonplanar N-point correlation functions, in 
which one of the external legs is twisted. Denoting the twisted vertex as the 
$N$-th, the relevant Feynman diagram is given in Fig.(\ref{diagram}):
\begin{figure}[htb]
   \vspace{0cm}
   \epsfysize=6cm
   \centerline{ \epsffile{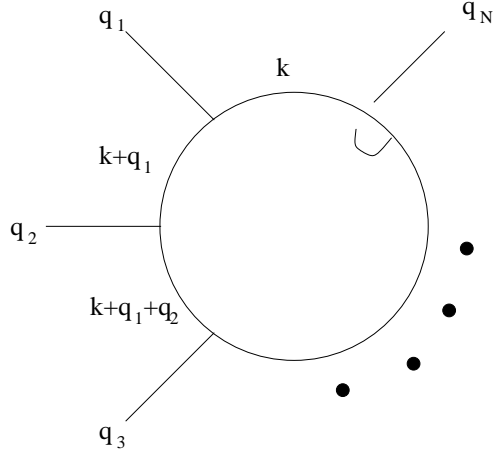} }
\caption{\sl Feynman diagram of one-loop, 
nonplanar N-point correlation function with N-th interaction vertex twisted.} 
\label{diagram}
\end{figure}
Evaulating Fig.(\ref{diagram}) explicitly, one obtains
the nonplanar one-loop correlation functions of the form Eq.(\ref{form}).
The Feynman loop integral ${\cal A}$ is independent of the noncommutativity
and hence has the same form as in commutative counterpart.  
In the parametrization of the internal and the external momenta as in 
Fig.(\ref{diagram}), after Wick rotation to Euclidean spacetime, 
the loop integral is given by
\bea
 {\cal A}\left(q_1, \cdots, q_{\rm N} \right) =  
\int {{\d^d} k \over (2 \pi)^d} \prod_{\ell=0}^{\rm N-1}
{1 \over (k+ q_1 + \cdots + q_\ell)^2 + m^2 }.
\label{1twist}
\eea
Introduce the Schwinger-Feynman parametrization for each propagator. This 
leads to
\bea
{\cal A} &=& \int {{\d^d} k \over (2 \pi)^d} 
\int_0^\infty \prod_{\ell=1}^{\rm N} \d \alpha_\ell \, 
\exp \left[-\alpha (m^2 + k^2)
  -2\left(\sum_{m=1}^{{\rm N}-1}\alpha_{m+1}
(q_1 + \cdots+ q_m )\right) \cdot k + {\cal O}(q^2) 
\right],
\nonumber
\eea
where $\alpha :=\sum_{\ell=1}^{\rm N} \alpha_\ell$ serves as the moduli
measuring perimeter of the one-loop graph. As we are mainly
concerned with the low-energy, large noncommutativity limit, 
Eqs.(\ref{limit1}, \ref{limit2}), we will drop terms of order ${\cal O}(q^2)$
in what follows.  

The Moyal phase factor, ${\cal B}$, is extractible from the ${\bf U}$ and
${\bf T}$ factors for untwisted and twisted interaction vertices, 
respectively. One finds, in counter-clockwise convention for Wick
contraction,
\bea
{\cal B}\left(q_1, \cdots, q_{\rm N} \right) = 
\exp \left[ i k \w \left( \sum_{\ell=1}^{\rm N-1} 
q_\ell \right)\right]
        \exp \left[ +{i \over 2} \sum_{\ell < m}^{\rm N-1} q_\ell \w q_m \right].
\nonumber
\eea
The integrand in ${\cal A}$ is simplified once the loop momentum variable 
is shifted as: 
\bea
k^\mu \longrightarrow k^\mu -{1 \over \alpha} 
\l[ \sum_{m=1}^{\rm N-1} \alpha_{m+1} \l( q_1 +\cdots+q_m \r) \r].
\nonumber
\eea
Change the the Schwinger-Feynman moduli parameters into those for
the ordered interaction vertices around the circumference of the 
one-loop diagram:
\bea
\alpha_\ell \longrightarrow \alpha (x_{\ell-1}-x_\ell)  
   \qquad {\rm where} \qquad
\left( \, 1> x_1 > x_2 > \cdots > x_{\rm N-1} >0 \, \right).
\nonumber
\eea
Accordingly, ${\cal B}$ is multiplied by a moduli-dependent phase factor:
\bea
{\cal B}(q_1, \cdots, q_{\rm N})
\longrightarrow {\cal B} (q_1, \cdots, q_{\rm N})
\times \exp\left( -i \sum_{\ell<m}^{\rm N-1} (x_\ell - x_m )q_\ell \w q_m \r),
\nonumber 
\eea  
and hence mixes with ${\cal A}$ through the (N$-1)$ 
moduli-parameter integrations.

The loop momentum and overall moduli integrals can be calculated 
explicitly. The loop momentum integral yields: 
\bea
 \int {\d^d k  \over (2 \pi)^d}\, \exp \left(-\alpha k^2 + i q_{\rm N} \w k \right)  
      = \l( {1 \over 4 \pi \alpha }\r)^{{d \over 2}} 
  \exp \left(- { q_{\rm N} \circ q_{\rm N}  \over 4 \alpha} \right),
\eea
while the $\alpha$-moduli integral yields:
\bea
 \int_0^\infty \d \alpha \, \alpha^{{\rm N}-1-{d \over 2}} 
    \exp\left[-m^2 \alpha - { q_{\rm N} \circ q_{\rm N}  \over 4 \alpha} 
     \right]   
   = 2^{{d \over 2}-\rm N+1} 
     \l( { {q}_{\rm N} \circ {q}_{\rm N} \over m^2}  
\r)^{{\rm N\over 2}
     -{d \over 4}}
     {\bf K}_{{d \over 2}-{\rm N}} \l( m \vert {q}_{\rm N} 
\circ {q}_{\rm N} \vert^{1 \over 2} \r).
\label{diver}
\eea
The remaining integrals over $\rm (N-1)$ moduli 
parameters are given by: 
\bea
\widetilde{{\cal B}}^{\rm (N)}_{\rm total} 
&=& \exp\left({i \over 2} \sum_{\ell<m}^{\rm N-1} q_\ell \w q_m \right)
        \int_0^1 \d x_1 \int_0^{x_1} \d x_2 \cdots \int_0^{x_{\rm 
N-2}} \d x_{\rm N-1} 
        \exp\l( -i \sum_{\ell<m}^{\rm N-1} (x_\ell - x_m ) q_\ell \w q_m \r)
\nonumber \\
&+& ({\rm permutations}).
\nonumber
\eea
The (permutations) part refers to (N$-1)$! diagrams of the same sort, 
differing from one another by permutation of the $\rm (N-1)$ untwisted 
interaction 
vertices.
Half of these diagrams are  Hermitian conjugates of the other, corresponding
to topological reversal between the inner and the outer sides. The complete
nonplanar N-point correlation function with a single twisted vertex is 
given by all possible cyclic permutation of the twisted interaction
vertex with the (N$-1)$ untwisted ones. 

For N=3 correlation function, two diagrams of ordering (1-2-[3]) and 
(2-1-[3]), in which [3] refers to the twisted vertex, contribute. The result 
is:
\bea
\widetilde{\cal B}^{(3)}_{\rm total} &=& 
e^{{i \over 2} q_1 \w q_2} \int_0^1 \d x \int_0^x \d y 
       \exp \Big( -i (x-y) q_1 \w q_2 \Big) \quad + \quad
( 1 \longleftrightarrow 2) 
\nonumber \\
 &=& \sin \left({q_1 \w q_2 \over 2 } \right) \left\slash
  \left( {q_1 \w q_2 \over 2 } \right) \right. ,
\eea
yielding precisely the generalized star product $\star_2$. 
Note that, in deriving the result, higher-powers of the momenta cancel out 
each other \cite{factorization}, leaving only the $\star_2$ part. 

For N=4 correlation function, there are $3!$ diagrams. Three diagrams with 
ordering (1-2-3-[4]), (1-2-[4]-3), and (1-[4]-2-3) contribute, 
while the other three diagrams (1-[4]-3-2), (1-3-[4]-2), and (1-3-2-[4]) are 
Hermitian conjugates, respectively. Summing them up, one finds:
\bea
\widetilde{\cal B}^{(4)}_{\rm total} &=& 
e^{{i \over 2} ( q_1 \w q_2 + q_1 \w q_3 + q_2 \w q_3 )} 
          \int \cdots \int e^{-i [(x_1 - x_2) q_1 \w q_2 + 
          (x_1 - x_3) q_1 \w q_3 +
            (x_2 - x_3) q_2 \w q_3 ]}  \nonumber \\
&+& e^{{i \over 2} ( q_3 \w q_1 + q_3 \w q_1 + q_1 \w q_2 )} \int \cdots \int
e^{-i [(x_1 - x_2) q_3 \w q_1 + (x_1 - x_3) q_3 \w q_2 +  
            (x_2 - x_3) q_1 \w q_2 ]}  \nonumber \\
&+& e^{{i \over 2} ( q_2 \w q_3 + q_2 \w q_1 + q_3 \w q_1 )} 
          \int \cdots \int
e^{-i [(x_1 - x_2) q_2 \w q_3 + (x_1 - x_3) q_2 \w q_1 +  
            (x_2 - x_3) q_3 \w q_1 ]}  
\nonumber \\
&+& ({\rm h. c.})  \nonumber \\
      &=&  \left( \sin {{q_2 \w q_3} \over 2} \left\slash 
{ {(q_1 + q_2) \w q_3} \over 2} \right) \right.
\left(\sin { q_1 \w (q_2+ q_3 ) \over 2} \left\slash 
{q_1 \w (q_2+ q_3 ) \over 2} \right. \right) 
   +  ( 1 \longleftrightarrow 2),
\nonumber
\eea
yielding precisely the generalized $\star_3$ product. 
  
The N=4 correlation function contains another type of nonplanar diagram, 
in which two interaction vertices are twisted. It consists of six diagrams
of distinct permutations, among which half are Hermitian conjugates of 
the others. Adding them up, one easily finds
\bea
 {\bf \Gamma}^{(4)} = \left( {\lambda \over 2} \right)^4 (2 \pi)^d
\widetilde{\cal K}_{4 - {d \over 2}}
\Big( (q_3 + q_4) \circ (q_3 + q_4) \Big) 
   \left( { \sin { q_1 \w q_2 \over 2}\over   { q_1 \w q_2 \over 2}}\right)
   \left({\sin{q_3 \w q_4 \over 2} \over {q_3 \w q_4 \over 2}}\right),
\nonumber
\eea
viz. product of two $\star_2$'s. 

Hence, up to ${\cal O}(\lambda^4)$, after Wick-rotation back to Minkowski
spacetime, the non-planar part of the one-loop effective action is given by: 
\bea
 \Gamma_{\rm np}[\Phi] &=& {1 \over 3!}
\left( - {\lambda \over 4} \right)^3  
  \int \d^d x  \left( 3 [\Phi(x) \Phi(x)]_{\star_2} 
   \, {\cal K}_{3 - {d \over 2}} \l(-\partial_x \circ \partial_x \r) 
   \, \Phi(x)  \right) \nonumber\\
 &+& {1 \over 4!}
\left( - {\lambda \over 4} \right)^4  \int \d^dx
  \l( 3 \, [\Phi(x) \Phi(x)]_{\star_2} \, {\cal K}_{4 - {d \over 2}}
 \Big(-(\partial_{x_1} +\partial_{x_2}) \circ 
 (\partial_{x_1} +\partial_{x_2}) \Big)  [\Phi(x_1)\Phi (x_2)]_{\star_2} 
\right.  
  \nonumber \\
 & & \hskip2.7cm  \left. + 4 \left[ \Phi(x) \Phi(x) \Phi(x) \right]_{\star_3} 
 \, {\cal K}_{4 - {d \over 2}}  \l(- \partial_x  \circ \partial_x  \r) 
  \Phi(x) \r) \nonumber \\
&+& \cdots,
\label{34loop}
\eea
where, for the second line, $x_1, x_2 \rightarrow x$ is assumed in the end.
The final expression is in complete agreement with power-series expansion
of Eq.(\ref{final}), as given in Eq.(\ref{seriesform}).

\subsection{Further Remarks} 
We close this section with two remarks concerning our result. 
First, contribution of planar part to the one-loop effective action can 
be deduced straightforwardly by replacing, in Eqs.(
\ref{final}, \ref{seriesform}), $\vert \theta \cdot {\bf k} \vert$ 
into $2 / \Lambda_{\rm UV}$, where $\Lambda_{\rm UV}$ is the ultraviolet
cutoff, and all the generalized $\star_n$ products into Moyal's 
$\star$-product.    
In the limit $\Lambda_{\rm UV} \rightarrow \infty$, utilizing Taylor
expansion of the modified Bessel function ${\bf K}_n$, 
one easily recognizes that
the planar part of the effective action reduces to `Coleman-Weinberg'-type 
potential and appears not to be expressible in terms of the open Wilson line 
operators, even including those of {\sl zero} momentum. 

Second, one might consider the results in this work trifling as, at 
low-energy, 
large noncommutativity limit Eqs.(\ref{limit1}, \ref{limit2}), the kernel 
$\widetilde{\cal K}_{-d/2} \sim \exp ( - m \vert \theta \cdot {\bf k} \vert)$
is exponentially suppressed. Quite to the contrary, we actually believe that 
the exponential suppression indicates a sort of {\sl holography} in terms of 
a gravity theory. Indeed, in the context 
of noncommutative D3-branes, as shown in \cite{rey2}, Green's functions of 
supergravity fields exhibit precisely the same, exponentially suppressed 
propagation in the region where the noncommutativity is important, viz. the 
dipole-dominated, `UV-IR proportionality' region. We view this as an evidence 
that, in noncommutative field theories, the dipoles are provided by open 
Wilson line operators and that their dynamics describes, via 
holography, gravitational interactions. The latter then defines \underline{
an effective field theory of the dipoles}. 

\section*{Acknowledgement}
SJR would like to thank H. Liu for stimulating correspondences, and L. 
Susskind for useful remarks. 


\end{document}